\documentclass[%
 aip,
 amsmath,amssymb,
 reprint,%
]{revtex4-1}

\usepackage{graphicx}
\usepackage{dcolumn}
\usepackage{bm}

\usepackage[utf8]{inputenc}
\usepackage[T1]{fontenc}
\usepackage{mathptmx}
\usepackage{etoolbox}
\usepackage{xcolor}
\usepackage{hyperref}

\makeatletter
\def\@email#1#2{%
 \endgroup
 \patchcmd{\titleblock@produce}
  {\frontmatter@RRAPformat}
  {\frontmatter@RRAPformat{\produce@RRAP{*#1\href{mailto:#2}{#2}}}\frontmatter@RRAPformat}
  {}{}
}%
\makeatother
\begin{document}

\preprint{AIP/123-QED}

\title{Machine learning many-body potentials for charged colloids in primitive 1:1 electrolytes}

\author{Thijs ter Rele}
\affiliation{Soft Condensed Matter \& Biophysics, Debye Institute for Nanomaterials Science, Utrecht University, Princetonplein 1, 3584 CC Utrecht, Netherlands}
\email{t.r.terrele@uu.nl}
\author{Gerardo Campos-Villalobos}
\affiliation{Soft Condensed Matter \& Biophysics, Debye Institute for Nanomaterials Science, Utrecht University, Princetonplein 1, 3584 CC Utrecht, Netherlands}
\affiliation{CNR-ISC and Department of Physics, Sapienza University of Rome, p.le A. Moro 2,  00185 Roma, Italy}
\author{Ren\'{e} van Roij}
\affiliation{Institute for Theoretical Physics, Utrecht University, Princetonplein 5, 3584 CC Utrecht, Netherlands}
\author{Marjolein Dijkstra}
\affiliation{Soft Condensed Matter \& Biophysics, Debye Institute for Nanomaterials Science, Utrecht University, Princetonplein 1, 3584 CC Utrecht, Netherlands}

\date{\today}

\begin{abstract}
Effective interactions between charged particles dispersed in an electrolyte are most commonly modeled using the Derjaguin-Landau-Verwey-Overbeek (DLVO) potential, where the ions in the suspension are coarse-grained out at mean-field level. However, several experiments point to shortcomings of this theory, as the distribution of ions surrounding colloids is governed by nontrivial correlations in regimes of strong Coulomb coupling (e.g. low temperature, low dielectric constant,  high ion valency, high surface charge).  
Insight can be gained by explicitly including the ions in simulations of these colloidal suspensions, even though direct simulations of dispersions of highly charged spheres are computationally demanding. To circumvent slow equilibration, we employ a machine-learning (ML) framework to generate density-dependent ML potentials that accurately describe the effective colloid interactions  at given system parameters. These ML potentials enable fast simulations and make large-scale simulations of charged colloids in suspension possible, opening the possibility for a systematic study of their phase behaviour, in particular gas-liquid and fluid-solid coexistence.
\end{abstract}

\maketitle

\section{Introduction}

The interactions between charged colloidal particles dispersed in liquids are of central interest in soft matter physics, as they govern the properties and phase behavior of charge-stabilized colloidal suspensions. These systems, composed of Brownian charged particles, are typically described by a combination of electrostatics and statistical physics using the Poisson-Boltzmann framework, which underlies the Derjaguin-Landau-Verwey-Overbeek (DLVO) theory.\cite{Derjaguin1941, Verwey1947} 
The pairwise DLVO potential provides an effective coarse-grained "colloids-only" description in which the ion  degrees of freedom are integrated out. It has become a cornerstone of colloid science and has proven highly effective in modeling the interactions and phase behavior of charged colloidal suspensions.\cite{Hunter2000,Israelachvili2011, Trefalt2017}  
The DLVO potential accounts for the electrostatic interactions between colloids suspended in an electrolyte, where each colloid is surrounded by an electric double layer (EDL) consisting of a cloud of mobile ions of opposite charge to the colloid   that screen its charge.\cite{Hansen2000}

When the refractive index of the solvent is matched to that of the colloidal particles, 
the attractive part of the DLVO potential, arising from  Van der Waals forces due to dipole-dipole fluctuations, can be neglected. Under these conditions, the interaction between two colloidal particles with the same charge sign reduces to  a purely repulsive electrostatic potential. 
It was therefore surprising when a series of experiments reported attractive interactions between like-charged colloids in systems  of highly charged, micron-scale particles at low-salt concentrations,\cite{Kepler1994, Ito1994, Tata1997, Larsen1997, Gomez2009, Wang2024} challenging the universal validity of DLVO theory. It should be noted, however, that the results of some of these experiments remain controversial, due to issues related to reproducibility \cite{Palberg1994_tata} and potential experimental artifacts.\cite{Squires2000_brenner} 
This low-salt regime is characterized by long screening lengths, which increases the importance of many-body interactions beyond simple pairwise interactions, as the  larger electric double layers more frequently overlap with multiple neighboring EDLs. A second characteristic of this regime is that, as the colloid concentration increases, the ionic strength, and thus the effective Debye length, is no longer determined predominantly by the concentration of the background electrolyte, but rather by the higher concentration of counterions. Various studies based on Poisson-Boltzmann calculations have proposed that these two effects are the leading causes behind like-charge attractions---specifically, gas-liquid and gas-crystal phase separation---in low-salt colloidal suspensions.\cite{vanRoij1997, vanRoij1999, Zoetekouw2006, Warren2000, Russ2002} 
A second cause of like-charged attractive interactions lies in the strongly coupled electrostatic regime, which is characterized by low temperature, low dielectric constant, and high ionic valency.\cite{Kjellander1988, Allahyarov1998, Naji2005, Punkkinen2008} A defining feature of this regime is the binding of point-like ions onto highly charged surfaces, forming a quasi-two-dimensional layer rather than the diffuse three-dimensional distribution characteristic of the Poisson-Boltzmann electric double layer. When two such decorated surfaces approach each other, strong ion-ion correlations give rise to attractive interactions.

To elucidate the mechanism behind non-DLVO effects in electrostatic interactions, researchers have performed fine-grained simulations of charged colloids suspended in an electrolyte. Unlike DLVO theory,  which integrates out the ionic degrees of freedom, these primitive model (PM) simulations explicitly represent the ions while treating the solvent as a dielectric continuum. This explicit inclusion of ions in these simulations is of vital importance, as like-charge attractions are potentially a result of the buildup of counterions near colloidal surfaces. 
Remarkably, these primitive model simulations have reproduced some of the anomalous like-charge attraction phenomena observed experimentally---features that are absent in simulations relying solely on DLVO-based colloidal interactions. These features include clustering of like-charged colloidal particles in  electrolytes with large charge- and size-asymmetries,\cite{Linse1999, Rescic2001}  broad gas-liquid and gas-crystal coexistence regions,\cite{Hynninen2007, Hynninen2009}  and attractive two-body potentials between colloids  in the strong Coulomb coupling regime.\cite{Wu1998, Wu1999, Lin2021} 
There is, however, a downside to the primitive model simulations described above: they are very computationally intensive, particularly for systems with high charge asymmetries. To ensure charge neutrality in the simulation box, a large number of charge-neutralizing counterions must be added. This large number of particles, combined with the long-range nature of Coulombic interactions, significantly slows down primitive model simulations and limits them to systems with low colloidal charge valency. For similar reasons, simulations at high-salt concentrations have often been avoided. However, studying the anomalous like-charge attractions observed experimentally requires simulations  at the high colloid charge valencies encountered in those experiments.

Several approaches have been proposed to speed up simulations with large charge asymmetries. One approach involves confining particle positions to a grid, allowing Coulombic interactions to be precomputed at the start of the simulation.\cite{Hynninen2005} 
Another strategy is to coarse-grain the ion densities surrounding the colloids using  classical density functional theory.\cite{Lowen1992, Lowen1993, Fushiki1992} Simulations employing this approach within a Car-Parrinello Molecular Dynamics framework for classical systems have achieved some success in capturing the structure of highly charged spherical colloids in suspension. However, deviations from primitive model simulations were observed at small colloid separations, indicating limitations in accurately describing short-range interactions.\cite{Lowen1997, Tehver1999}

In this paper, we also coarse-grain the ions to significantly reduce simulation times, but we achieve this by employing a machine learning (ML) framework to generate effective many-body colloidal potentials. While ML approaches have been successfully used to accelerate simulations of uncharged systems,\cite{Deringer2019, CamposVillalobos2021, Nguyen2022, Argun2024}  the present work is the first to apply machine-learned effective potentials to charged colloids.
To construct these ML colloidal potentials, we adopt a linear regression scheme previously used to develop effective potentials for sterically stabilized colloids with non-adsorbing polymers \cite{CamposVillalobos2021} and for ligand-stabilized colloids.\cite{Giunta2023}  
Here, we train the ML potentials using data from primitive model simulations, resulting in effective potentials that accurately capture the effective interactions between charged colloids in the presence of ions, in addition to bare-colloid interactions. 
In subsequent ``colloids-only'' simulations using these ML potentials, the ions are effectively integrated out.
Moreover, in this representation,    the simulations do not include any explicit charges, since the colloids interact through the effective coarse-grained ML potential, which eliminates the need to account for  long-range interactions. 
These two factors bring about a significant increase in computational efficiency
--at least twenty times faster for colloidal systems--, enabling studies at higher colloid charges and salt concentrations.

In Section \ref{sec:model-method}, we introduce the framework for generating effective ML potentials. In Section \ref{sec:colsys}, we  apply these ML potentials in simulations to evaluate their performance under realistic system parameters corresponding to  an experimental colloidal suspension in a low-polar solvent. In Section \ref{sec:Z50}, we apply the same framework  to describe interactions between colloids in a low-temperature, salt-free suspension. Finally, in Section \ref{sec:elsys}, we test the limits of the framework by using our ML method to generate effective potentials between cations in an aqueous electrolyte. 

\section{Model and Methods}\label{sec:model-method}

\subsection{Primitive Model simulations}
\begin{figure}[ht]
  \centering
  \hspace*{-0.0cm}
    \includegraphics[width=\linewidth]{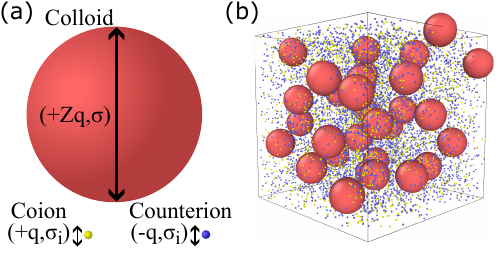}
  \centering
  \caption{(a) Primitive model of a colloidal dispersion in a 1:1 electrolyte, represented as a three-component mixture of charged colloids, counterions, and coions with charges $Zq$, $-q$, and $+q$,  and diameters $\sigma$, $\sigma_i$, and  $\sigma_i$, respectively.  (b) Typical configuration from a primitive model simulation with $N_c = 32$ colloidal particles of valency $Z=90$, $N_+=2551$ coions, $N_-=5431$ counterions,  and size ratio  $\sigma_i/\sigma=0.05$.  This snapshot is representative of the simulations used to generate the training data.}
  \label{fig:illustrate_size}
\end{figure}

The system of interest in this study is a three-dimensional primitive model (PM) of a colloidal dispersion in a 1:1 electrolyte. This colloidal suspension is viewed as a three-component mixture of charged colloids, monovalent counterions, and monovalent coions, while the solvent is modeled as a structureless dielectric continuum. The colloidal particles have an effective hard-core diameter $\sigma_c$ and a total positive charge $Q_c q$ uniformly distributed over their surfaces, with valency $Q_c\geq 1$ and elementary charge $q$. The monovalent counterions are anions with diameter $\sigma_-$ and charge $-q$, and the monovalent coions are cations of diameter $\sigma_+$ and charge $+q$. These three species are illustrated in Fig. \ref{fig:illustrate_size}, where we ease the notation and use the short-hand  $\sigma_c=\sigma$ for the colloid diameter, $Q_c = Z$ for the colloid charge number, and $\sigma_+=\sigma_-=\sigma_{i}$ for the ion diameter that is assumed to be equal for the cations and the anions.  

Interactions between these particles are modeled as the sum of a pseudo-hard sphere repulsion, described by the Weeks-Chandler-Andersen potential, and an electrostatic interaction represented by the Coulomb potential. The Weeks-Chandler-Andersen (WCA) potential between particle $k$ and particle $l$ at center-to-center distance $r_{kl}$ is given by
\begin{equation}
    \hspace{-2mm} U_{WCA}(r_{kl}) =\begin{cases}4\overline{\varepsilon}\left[\left(\frac{\sigma_{kl}}{r_{kl}}\right)^{12} \hspace{-2mm} - \left(\frac{\sigma_{kl}}{r_{kl}}\right)^{6} \right] + \overline{\varepsilon},& \  \hspace{-2mm} r_{kl} \leq 2^{1/6} \sigma _{kl};
        \\ 0, & \  \hspace{-2mm} r_{kl} > 2^{1/6} \sigma_{kl}, 
    \end{cases}
\end{equation}
where $\sigma_{kl} = (\sigma_k + \sigma_l)/2$ is their (quasi-) contact distance and $\beta \overline{\varepsilon} = 40$, with $\beta = 1/k_B T$, $T$ the temperature, and $k_B$ the Boltzmann constant. The Coulomb potential between pairs of particles reads 
\begin{equation}
    \beta U_{C}(r_{kl}) = \frac{Q_k Q_l \lambda_B}{r_{kl}} ,
\end{equation}
where $Q_k$ denotes the charge number (the valency) of particle $k$, and where the Bjerrum length $\lambda_B = \beta q^2/(4\pi \epsilon)$ determines the strength of the electrostatic interactions in a structureless dielectric medium of permittivity $\epsilon$ at temperature $T$. 

In a fixed volume $V$, we denote the number of colloids by $N_c$ and the number of counter- and co-ions by $N_-$ and $N_+$, respectively. Global charge neutrality imposes $Q_c N_c + N_{+}=N_{-}$ at all times, both in the canonical ensemble, where the particle numbers of all species are fixed, and in the semi-grand ensemble, where only $N_c$ is fixed while $N_+$ and $N_-$ can fluctuate at a fixed chemical potential $\mu$ of the salt. 

The thermodynamic equilibrium properties of this primitive model are governed by six dimensionless parameters, which reduce to five by considering equal ion diameters $\sigma_+=\sigma_-\equiv \sigma_i$. Using the shorthand notation $N_c=N$, the first four key parameters are (i) the colloid packing fraction $\eta=(\pi \sigma^3/6)N/V$, (ii) the ion-to-colloid size ratio $\sigma_i/\sigma$, (iii) the colloid-to-ion charge asymmetry or colloid valency $Z$, and (iv) the size ratio $\sigma/\lambda_B$, which plays the role of an effective (dimensionless) temperature. 
In addition, we require a fifth dimensionless parameter to characterize the salt concentration, which can be expressed in various ways, such as the ion packing fraction $(\pi \sigma_i^3/6)(N_++N_-)/V$, the ion activity $\exp(\beta\mu)$, or the ion number ratio $N_+/N_-$, which varies from 0 in the salt-free case to unity in salt-dominated systems. In this work, we opt for the $\beta\mu$-dependent combination 
$\kappa\sigma$ as the fifth dimensionless parameter, where $\kappa$ is the inverse Debye screening length of the colloid-free salt reservoir in osmotic equilibrium with the colloidal suspension. This implies that the chemical potential of the salt in the reservoir is equal to that in the suspension. At a fixed salt (ion) chemical potential $\mu$, or more precisely $\beta\mu$, the inverse Debye screening length in the salt reservoir is given by 
$\kappa= \sqrt{4\pi\lambda_B c_s}$, where $c_s = (n_+ + n_-)$ is the salt concentration in the salt reservoir, consisting of counterion concentration $n_{+}$ and coion concentration $n_{-}$. In the reservoir, these concentrations are equal, so $n_+ = n_-$. It is important to note that this reservoir concentration differs from the total (thermally averaged) salt concentration in the suspension $\big(N_+(\mu,\eta)+N_-(\mu,\eta)\big)/V$, or the added salt concentration $2N_+(\mu,\eta)/V$. However,  these quantities coincide in the colloid-dilute limit  $\eta\rightarrow 0$ or in the high-salinity limit $N_+/N_-\rightarrow 1$.

\subsection{Training Data Generation} \label{sec:traingen}
The training data required to construct an effective machine-learned (ML) colloidal potential is obtained from molecular dynamics (MD) simulations using the primitive model. In particular, the ML potential is trained on the average PM force ${\bf F}^{PM}_i$ on colloid $i=1,\dots,N$ in specific colloid configurations $\{{\bf R}\}$, where the averaging is performed over the ion degrees of freedom of the primitive model.  Each training set consists of configurations sampled at colloid packing fractions within the range $\eta \in [0.001, 0.45]$, while keeping  $Z$, $\sigma_i/\sigma$, $\sigma/\lambda_B$, and $\beta\mu$ fixed. 
Our goal is to generate an ML  potential that can be used to simulate a system characterized by $Z$, $\sigma_i/\sigma$, $\sigma/\lambda_B$, and $\beta\mu$ across a range of packing fractions, thereby enabling the study of not only single bulk phases at varying densities but also phase-separated systems with  different densities. It should be noted, however, that a separate ML potential must be trained for each distinct set of system parameters ($Z$, $\sigma_i/\sigma$, $\sigma/\lambda_B$, $\beta\mu$).

To initialize a simulation with given particle sizes, colloid valency, and temperature, we begin by randomly placing $N$ colloids and $N_-=ZN$ charge-neutralizing counterions in a cubic box of length $L =  V^{1/3}$, while preventing hard-sphere overlaps between particles and where $V$ is chosen such that the desired packing fraction $\eta = (\pi \sigma^3/6)N/L^3$ is obtained. The data-generating primitive model simulations involve a fixed number of $N$ colloids and are performed using the LAMMPS software package.\cite{LAMMPS}  The equations of motion are integrated using  the Verlet algorithm, and the Nos\'{e}-Hoover thermostat is employed to maintain a constant  temperature of $k_B T /\epsilon = 1.0$, with a relaxation time of $0.05 \tau_{MD}$, where  $\tau_{MD}  =\sqrt{m \sigma^2/(k_B T)}$.\cite{FrenkelenSmit} The long-range Coulombic interactions  for distances $R > 2.5\sigma$ are computed using the particle-particle-particle-mesh (PPPM) Ewald summation method, with a target relative force error of $10^{-4} k_BT/\sigma$.\cite{Hockney1988}   Periodic boundary conditions are applied to mimic bulk behavior. Below, we distinguish between the salt-free limit $\exp{[\beta\mu]} \rightarrow 0$, which we will also refer to as the asymmetric electrolyte because only colloids and counterions are present, and the case with added salt, where pairs of counterions and coions are inserted into or removed from the simulation box using a Grand Canonical Monte Carlo scheme controlled by the salt chemical potential $\beta \mu$.\cite{FrenkelenSmit} 

Within this Grand Canonical Monte Carlo scheme, ions are inserted as charge-neutral counterion-coion pairs that are initially placed at a distance of $0.1 \sigma$ so they lie close to one another but do not interact through the WCA-potential, as all ions have a diameter $\sigma_i = 0.05\sigma$ in the simulations performed in this paper. The acceptance rate for inserting or removing ion pairs depends on two factors. The first is the difference in potential energy $\mathcal{U}$ of adding or removing an ion pair. If the energy cost of insertion or removal is low (or even negative), it is more likely to be accepted. The second factor is the chemical potential $\mu$ of the salt pair.  Within the implemented Grand Canonical Monte Carlo framework, the insertion or removal of an ion pair  is accepted with probabilities
\begin{align}
\text{acc}(N_s\rightarrow N_s+2) = &\text{min}\Big[1,
\\ \nonumber
\hspace{-5mm}\frac{V}{(N_s+2)\Lambda^3 }& \exp{\big[\beta(\mu - \mathcal{U}(N_s+2) + \mathcal{U}(N_s))\big]} \Big]; \\
\text{acc}(N_s \rightarrow N_s-2) = &\text{min}\Big[1, \\ \nonumber & \hspace{0mm}
\hspace{-15mm}\frac{N_s\Lambda^3}{V}  \exp{\big[-\beta(\mu + \mathcal{U}(N_s-2) -\mathcal{U}(N_s))\big]} \Big], 
\end{align}
where $\mathcal{U}(N_s)$ represents the potential energy associated with a configuration with $N$ colloids and $N_s=N_++N_-$ salt ions,\cite{FrenkelenSmit}  and where we set the thermal wavelength $\Lambda = \sigma$  equal to the colloid diameter, for convenience.

In the case of a salt-free colloidal system with monovalent counterions, the primitive model system is first simulated for 20000 MD steps with a timestep of $\Delta t = 0.0005 \tau_{MD}$, where $\tau_{MD} = \sqrt{m \sigma^2/(k_B T)}$, with $m$ the particle mass, assumed to be the same for all species.\footnote{While assuming equal mass for all species is unrealistic for the dynamics, it does not affect the equilibrium properties of interest in this case.}  After this initial run, the positions of the colloids are fixed, and the simulation proceeds with 400000 MD steps during which only the counterions are allowed to move. The force acting on each colloid is recorded every 200 time steps, from which the ion-averaged PM forces ${\bf F}^{PM}_i$ are determined for the specific configuration.

For primitive model simulations with added salt, the procedure is as follows: after placing $N$ colloids and $ZN$ counterions in the box, the system of colloids and counterions is evolved for 10000 MD steps using a time step of $\Delta t =  0.0005 \tau_{MD}$. This initial simulation is followed by 25000  Grand Canonical Monte Carlo moves on the salt ions at the specified salt chemical potential $\beta \mu$, involving both insertion and removal of counterion-coion pairs.\cite{FrenkelenSmit}  After this, another 10000 MD steps are performed on the three-component primitive model, while every 200 steps 100 Grand Canonical Monte Carlo moves are performed. After this procedure, the colloid configuration, at given $\eta$ and $\beta\mu$, is fixed, however, the ions can still move during the subsequent 40000 MD steps. In the final stage, no Grand Canonical Monte Carlo moves are performed anymore, such that the ion numbers $N_+$ and $N_-$ stay constant, with the colloids still fixed in place. During this stage, the canonical system is simulated for 800000 MD steps, with the force on each colloid recorded every 200 time steps, from which the ion-averaged PM forces ${\bf F}^{PM}_i$ are determined for colloid $i=1,\dots, N$, for this particular colloid configuration. Simulations in which Grand Canonical Monte Carlo moves were performed while average colloid forces were determined resulted in similar results to those from the above procedure.

This procedure is repeated for $M$ different colloid configurations for both the salt-free limit and the added-salt case. For the given system parameters of the primitive model, this results in a set of $3N\times M$ Cartesian vector components of the ion-averaged forces acting on the colloids,
which, together with the $M$ stored colloid configurations $\{\bf{ R}\}$ are used as input data for the force-matching linear regression algorithm that we discuss now. 

\subsection{Machine Learning Procedure}\label{sec:ML-procedure}
We proceed with an ML procedure to construct a ML potential energy function $U^{ML}(\{\bf{ R}\})$ that depends on the colloid coordinates and represents a free energy (or grand potential). The scheme is designed such that the ML force $-\nabla_iU^{ML}(\{{\bf R}\})$ accurately matches the ion-averaged PM force ${\bf F}^{PM}_i$ acting on each colloid $i=1,\dots,N$ in configuration $\{{\bf R}\}$, as obtained from primitive model simulations, where the operator $\nabla_i$ denotes the gradient with respect to the center-of-mass position ${\bf R}_i$ of colloid $i$.  
Throughout the ML procedure, we assume that the potential energy function can be expressed as a sum of individual colloid contributions $U^{ML}(\{{\bf R}\})=\sum_{i=1}^N U^{ML}_i(\{{\bf R}\})$, where each energy contribution $U^{ML}_i$ is expressed in terms of a set of descriptors that characterize the local (rather than global) environment of colloid $i$ in configuration $\{{\bf R}\}$. In this work, we employ the atom-centered symmetry functions introduced by Behler and Parrinello \cite{Behler2007} to represent the local environment of each colloid, where the locality is characterised by a maximum cut-off distance $R_c$. We distinguish between two-body and three-body symmetry functions: the two-body terms characterize the environment of particle $i$  solely through distances $R_{ij}=|{\bf R}_j-{\bf R}_i|$ to nearby colloids $j$, while the three-body terms additionally incorporate angular dependencies. Specifically, to
characterize the local environment of colloid $i$ within a cutoff distance $R_c$ at the two-body level, we use symmetry functions
\begin{align}\label{G2}
    G^{(2)}(i) &= \sum_j e^{-\gamma(R_{ij} - R_s)^2} f_c(R_{ij}), 
\end{align}
where the sum runs over all neighboring particles $j$ and where $\gamma$ and $R_s$ are optimization parameters (see below) that control the width and center of the Gaussians with respect to $R_{ij}$, respectively. Here $f_c(R_{ij})$ is a cut-off function given by  
\begin{equation}
    f_c(R_{ij}) =  \begin{cases}
\tanh^3{\left(1 - R_{ij}/R_c\right)} & \text{for } R_{ij} \leq R_c; \\
0 &\text{for } R_{ij} > R_c,
\end{cases}
\end{equation}
which decays smoothly to zero in both value and slope at the cut-off distance $R_c$, ensuring smooth behavior at the boundary. At the three-body level, the local environment of particle $i$ is characterised by symmetry functions with distance and angular dependencies given by 
\begin{align}\label{G3}
    G^{(3)}(i) = & 2^{1-\xi} \sum_{j, k\neq i} \left(1 + \lambda \cos{\theta_{ijk}} \right)^{\xi} \nonumber
    \\ & e^{-\gamma \left(R_{ij}^2 + R_{ik}^2 + R_{jk}^2\right)} f_c(R_{ij}) f_c(R_{jk}) f_c(R_{ik}),
\end{align} where the sum runs over all distinct pairs 
$j,k\neq i$ within the cut-off distance $R_c$ to particle $i$ and $\theta_{ijk}$ is the angle between the vectors ${\bf R}_{ij}={\bf R}_j-{\bf R}_i$ and ${\bf R}_{ik}={\bf R}_k-{\bf R}_i$. The optimisation parameters $\gamma$, 
and $\lambda$ control the radial and angular resolution, respectively, and the parameter $\xi$ can only take values of $+1$ or $-1$.

With the functional forms Eqs.(\ref{G2}) and (\ref{G3}) of the descriptors of the local environment of each particle $i$ at hand, we now proceed by generating a large but manageable pool of $7\times(11+6\times2)= 161$ candidate descriptors, characterised by 7 values of $\gamma \in \{0.01, 0.1, 1, 2, 4, 8, 16\} \sigma^{-2}$, 11 values 
$R_s \in \{0.0, 0.1, 0.2, 0.3, 0.4, 0.5, 0.6, 0.7, 0.8, 0.9, 1.0\} \sigma$,  
6 values for $\lambda \in \{1, 2, 4, 8, 16, 32\}$, and 2 values
$\xi \in \{1, -1\}$. Throughout, we set the cut-off distance at $R_c=4.0\sigma$. 
From this initial pool, a subset of $D\ll161$ descriptors is selected using a feature selection method discussed below, aimed at identifying those parameters that best capture the local environment of the colloids. 
The selected symmetry functions, assumed known for now, are denoted by $G_k(j)$ for each colloid $j=1,\dots,N$ with $k=1,\dots,D$ indexing the selected symmetry functions.  
In our framework, which started from the assumption that the total potential energy $U^{ML}=\sum_{j=1}^NU^{ML}_j$ can be expressed as the sum of the potential energy $U^{ML}_j$ of the individual particles (with $j$ understood to be a dummy summation index that was called $i$ earlier), we now describe $U^{ML}_j$ as a weighted sum of symmetry functions  $G_k(j)$ with weights $\omega_k$ that are to be determined. Accordingly, we write \begin{equation}
    U^{ML}\left(\{\mathbf{R}\}\right) = \sum_{j=1}^{N} U^{ML}_j = \sum_{j=1}^{N} \sum_{k=1}^D \omega_k  G_k^{}(j), \label{eq:U_ML}
\end{equation}
where $j = 1,..., N$ labels the colloidal particles at positions $\{\mathbf{R}\}$ and $D$ is the total number of employed symmetry functions. As mentioned before, the effective ML force on particle $i$ is then obtained by taking the (negative) gradient of $U^{ML}$ with respect to position $\mathbf{R}_i$, which yields 
\begin{align}
    \mathbf{F}_{i}^{ML}  &=  -\nabla_i U^{ML} = - \sum_{j=1}^N \sum_{k=1}^{D} \omega_k \nabla_i G_k(j),
    \label{eq:SF_force} 
\end{align}
where we note that $i$ is \emph{not} a dummy summation index here. The gradients of the symmetry functions are given in Appendix \ref{sec:symgrad}. 
We also wish to note that by representing the ion-averaged PM forces $\mathbf{F}^{PM}_i$ by a simple linear model in terms of gradients of structural
descriptors, we also gain direct access to the many-body potential $U^{ML}\left(\{\mathbf{R}\}\right)$ of the coarse-grained model. Obtaining this potential is typically challenging, as it often requires  time-consuming thermodynamic integration methods.
The availability of $U^{ML}\left(\{\mathbf{R}\}\right)$ enables its direct use in Monte Carlo simulations. \cite{Giunta2023} 

We now gather the three Cartesian components of the ion-averaged PM forces $\mathbf{F}^{PM}_i$ for each colloid $i=1,\dots,N$ and for all $M$ configurations of our primitive model simulations into a single vector $\mathbf{f}^{PM}$ of dimension $3NM$. In other words, ${\bf f}^{PM}$ contains the three Cartesian force components for all $N$ particles across $M$ configurations. We now wish to approximate ${\bf F}^{PM}_i$ as closely as possible by ${\bf F}_i^{ML}$. Thus, we rewrite the condition ${\bf F}^{PM}_i\simeq{\bf F}_i^{ML}$ via Eq.~(\ref{eq:SF_force}) as the matrix equation
\begin{equation}
    \mathbf{f}^{PM} \simeq  \mathbf{X}\mathbf{\omega} \equiv \mathbf{f}^{ML},
\end{equation}
where $\mathbf{\omega}$ is a vector of dimension $D$, representing the weights of the symmetry functions, and $\mathbf{X}$ is a rectangular (rather than square) $3NM \times D$ matrix containing the derivatives $\partial_{\alpha, i} G_k(j)$ of all $D$ selected symmetry functions with respect to each particle position in each configuration. Here, $\partial_{\alpha, i}$ denotes the Cartesian component $\alpha = x,y,z$ of the gradient $\nabla_i$. To  determine the  weights of the symmetry functions in Eq.~(\ref{eq:SF_force}) we fit them to the ion-averaged PM colloid forces represented by ${\bf f}^{PM}$, using the equations of linear regression given by
\begin{equation}
    \mathbf{\omega} =  \left(\mathbf{X}^T \mathbf{X}\right)^{-1} \mathbf{X}^T \mathbf{f}^{PM},\label{eq:linear-regression}
\end{equation}
with $\mathbf{X}^T$ the transpose of $\mathbf{X}$.\cite{Bishop2006}  
Equation (\ref{eq:linear-regression}) yields the set of weights $\mathbf{\omega}$ that minimizes the sum of 
squared residuals $\left(\mathbf{f}^{PM} - \mathbf{X}\mathbf{\omega}\right)^2$, thereby also  minimizing the Root Mean Squared Error (RMSE), defined as  
\begin{equation}\label{RMSE}
    \text{RMSE} = \sqrt{\frac{\left(\mathbf{f}^{PM}- \mathbf{f}^{ML} \right)^2}{3NM}}
\end{equation}
which we will exploit now to actually determine the optimal subset of descriptors from the total pool of 161 symmetry functions.

To identify the optimal subset of descriptors, we use a feature selection method that minimizes the RMSE in a stepwise way by adding selected descriptors one by one.
We first identify which single symmetry function (out of the pool of 161) yields the lowest RMSE from Eq.~(\ref{RMSE}), where $\omega$ is determined via linear regression using Eqs.~(\ref{eq:linear-regression}), and $\mathbf{f}^{ML}$ using Eq. (\ref{eq:SF_force}) with $D=1$. The selected function is called $G_1(j)$. Next, we add a second descriptor to the first selected one, and determine the optimal weight via linear regression using $D=2$ in Eqs.~(\ref{eq:linear-regression}) and (\ref{eq:SF_force}). The descriptor that results in the smallest RMSE is then included in the set as $G_2(i)$, such that the set now contains two symmetry functions. This process is repeated for an increasing number of symmetry functions, incrementally adding the symmetry function that reduces the RMSE the most, until the RMSE converges to a minimum or decreases below a certain threshold, which then determines $D$ and the selected symmetry functions $G_k(i)$ for $k=1,\dots,D$. Typically, we have $D \simeq 20$.  

We also characterize the accuracy of the trained potentials using the coefficient of determination $R^2$, defined as
\begin{equation}
    R^2 = 1 - \frac{\left(\mathbf{f}^{ML} - \mathbf{f}^{PM}\right)^2}{\left( \bar{\mathbf{f}}^{PM} - \mathbf{f}^{PM} \right)^2},
\end{equation}
where $\bar{\mathbf{f}}^{PM}$ is a vector of dimension  $3NM$ in which each entry is equal to the mean of all components of  $\mathbf{f}^{PM}$, and thus $\bar{\mathbf{f}}^{PM} \cdot \mathbf{1} = \mathbf{f}^{PM} \cdot \mathbf{1}$, where $\mathbf{1}$ denotes the $3NM$-dimensional vector with all entries equal to one. 
The coefficient of determination quantifies how well  the ML forces reproduce the input forces from simulations, with $R^2 = 1$ indicating perfect agreement, corresponding to an RMSE of 0. 
 Once $R^2$ approaches $1$ and  the forces predicted by the ML potential closely match those from the training data  across their full range, we proceed by performing simulations using the ML potential.

\subsection{Coarse-Grained Simulations using ML potentials}
Once the coefficients $\omega_k$ are determined from Eq.~(\ref{eq:linear-regression}) for a given set of system parameters $\sigma_i/\sigma$, $\sigma/\lambda_B$, $Z$, and $\kappa\sigma$, for $D$ symmetry functions labeled $k=1,\dots,D$, the corresponding ML potential $U^{ML}(\{\bf R\})$ is explicitly known from Eq.~(\ref{eq:U_ML}). For all packing fractions within the training range, the resulting expression for $U^{ML}(\{\bf R\})$ can then be employed in Monte Carlo (MC) and MD simulations of effective colloids-only systems. These simulations contain many more colloidal particles than the 32 that were used in the primitive model simulations to determine $U^{ML}(\{\bf R\})$; typically we use hundreds or thousands of colloidal particles in the colloids-only systems.   
As with the primitive model simulations, the simulations employing ML potentials are MD simulations performed using the LAMMPS software package,\cite{LAMMPS}  where we make specific use of the HDNNP-package in implementing the many-body potentials. \cite{Singraber2019} 

\begin{figure}[h]
\hspace*{-1.cm}
\includegraphics[width=0.95\linewidth]{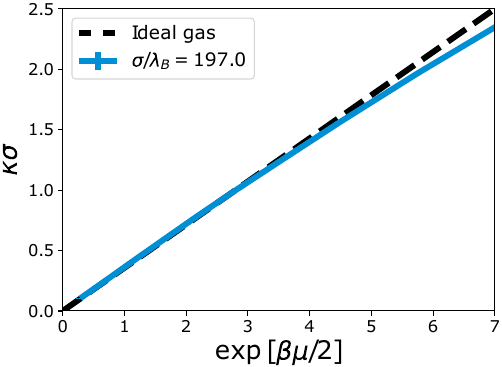}
  \caption{The inverse screening length $\kappa\sigma$ as a function of ion activity $\exp{\left[\beta \mu/2 \right]}$ with $\mu$ the chemical potential of salt pairs for a 1:1 electrolyte at temperature $\sigma_i/\lambda_B=9.85$ as obtained from Grand Canonical Monte Carlo simulations (blue line) along with the ideal gas prediction for comparison (dashed black line).  }
  \label{fig:bm_kapsig}
\end{figure}

\section{Colloidal System in Low-Polar Solvent}\label{sec:colsys}
As a proof of concept for generating ML potentials that accurately capture the effective colloidal system of the primitive model, 
we first apply our methodology to the well-characterised and well-understood experimental colloidal system studied by Royall \textit{et al.}.\cite{Royall2006}  This system consists of sterically stabilized poly-methyl methacrylate colloids with diameter $\sigma=2.16~\mu\text{m}$ and valency $Z=90$ in a solvent mixture of
cis-decalin and cyclohexyl bromide with a relative dielectric constant of 5.37 such that  
$\sigma/\lambda_B = 197$, which amounts to $Z\lambda_B / \sigma = 0.46$. Furthermore, we set the ion-to-colloid size ratio  $\sigma_i/\sigma=0.05$, and verify that the results are insensitive to further decreasing the ion size. 
To accurately reproduce the inverse screening length $\kappa \sigma = 1.56$  reported by Royall \textit{et al.}, we first determine the reservoir salt concentration as a function of the salt chemical potential $\beta \mu$ or ion activity $\exp{[\beta\mu/2]}$. 
To this end, we first perform MD simulations of the 1:1 electrolyte without colloids (effectively at $\eta = 0$) at given values of $\beta\mu$. Note that the effective (dimensionless) temperature of this electrolyte is given by $\sigma_i/\lambda_B=9.85$. After equilibration, we determine the average number of ions in the simulation box and use this to calculate the salt concentration $c_s$ and the corresponding inverse screening length $\kappa$. 
In Fig. \ref{fig:bm_kapsig}, we plot the inverse  screening length $\kappa \sigma$ against the ion activity $\exp{\left[\beta \mu/2\right]}$. In the dilute limit, where ions are non-interacting,  we expect a linear relationship between $\exp{\left[\beta \mu/2\right]}$ and $\kappa \sigma$, as the salt concentration is given by  $c_s = 2\exp(\beta\mu)/\Lambda^3$ within the ideal gas limit. As shown in Fig. \ref{fig:bm_kapsig}, our simulation results (blue line) show excellent agreement with the ideal gas prediction (dashed line)  at low values of $\exp{\left[\beta \mu/2\right]}$, corresponding to a dilute electrolyte. At higher ion activities, deviations appear: the simulation results yield a lower $\kappa \sigma$ than the ideal gas prediction due to the increasing importance of ion-ion interactions at higher concentrations. 
To accurately reproduce the experimental screening parameter $\kappa \sigma = 1.56$, we set the salt chemical potential to $\beta \mu = 3.0$ or ion activity $\exp{[\beta\mu/2]}=4.48$.

\begin{figure}[ht]
  \hspace*{-0.1cm}
  \includegraphics[width=0.98\linewidth]{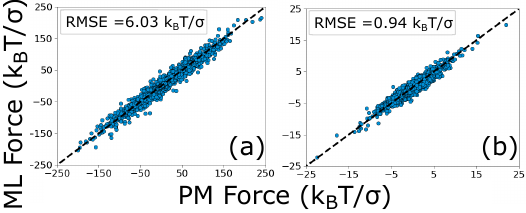}
  \caption{Parity plot comparing the Cartesian components of the effective many-body ML  forces $\mathbf{F}^{ML}_{i,\alpha}$ (in units of $k_B T/\sigma$)  predicted by the ML model with the corresponding PM forces $\mathbf{F}^{PM}_{i,\alpha}$ measured in primitive model simulations for the same configurations. The system consists of colloids with $Z =90$, effective temperature $\sigma/\lambda_B =  197.0$, inverse screening length $\kappa \sigma = 1.56$, and ion-to-colloid size ratio $\sigma_i/\sigma=0.05$.   The ML model was trained on configurations with packing fractions in the range (a) $\eta \in \lbrack 0.001, 0.45 \rbrack $ and (b) $\eta \in \lbrack 0.001, 0.1 \rbrack $, where the scale difference of the axes between (a) and (b) should be noted. }
  \label{fig:force_compare}
\end{figure}

We perform primitive model simulations on $N=32$ colloids at a salt chemical potential $\beta \mu = 3.0$ at 110 different packing fractions in the range $\eta \in \lbrack 0.001, 0.45 \rbrack$. For each packing fraction $\eta$, we collect 4 different colloid configurations $\{ \mathbf{R} \}$ and measure the ion-averaged forces ${\bf F}^{PM}_i$ acting on each colloid. Subsequently, we train a ML model using the ion-averaged forces obtained from $M=415$ colloid configurations spanning the entire packing fraction range $\eta \in \lbrack 0.001, 0.45 \rbrack$. The resulting ML potentials based on $D = 20$ symmetry functions have a coefficient of determination $R^2 = 0.953$  and Root Mean-Squared Error of $\text{RMSE} = 5.92 k_B T/\sigma$. In Fig. \ref{fig:force_compare}(a), we compare the coarse-grained many-body forces predicted by our ML model with those measured in the primitive model simulations. We find reasonable to good agreement between the ML-predicted forces and the effective many-body forces obtained from the primitive model. Fig. \ref{fig:force_compare}(b) shows the same comparison, however now with a training set that excludes the high packing fractions $\eta>0.1$ for reasons discussed below.

Due to the linearity of our ML model and the way it is trained, we also have direct access to the scalar coarse-grained many-body potential $U^{ML}(\{{\bf R}\})$. 
To gauge the accuracy of the effective ML potential $U^{ML}(\{{\bf R}\})$--- trained on the PM-generated force-vector ${\bf f}^{PM}$---we evaluate the effective ML potential for a system consisting of only two colloids $(N=2$) at positions ${\bf R}_1$ and ${\bf R}_2$, such that we can define the effective ML pair potential 
\begin{equation}\label{U2}
    U^{ML}_2(R_{12})\equiv U^{ML}({\bf R}_1,{\bf R}_2),
\end{equation}
where we set the single-body potential $U^{ML}({\bf R}_1)=0$. Likewise, we can define for $N=3$ colloidal particles the effective 3-body ML potential
\begin{eqnarray}
\label{U3}
U^{ML}_3(R_{12},R_{13},R_{23})&=&\nonumber \\
&&\hspace{-40mm}U^{ML}({\bf R}_1,{\bf R}_2, {\bf R}_3) -U^{ML}_2(R_{12})-U^{ML}_2(R_{13})-U^{ML}_2(R_{23}),
\end{eqnarray}
where we note that its dependence on the three scalar distances can also be rewritten in terms of two distances and an angular variable. The ML predictions $U^{ML}_2$ and $U^{ML}_3$, trained on primitive model simulations with $N=32$ colloidal particles, can also be compared to results from ion-averaged PM forces obtained from separate primitive model simulations with $N=2$ and $N=3$ colloids, respectively, at fixed positions in a cubic simulation box with sides of length $L = 10 \sigma$, equal to $15.6 \kappa^{-1}$ and corresponding to colloid packing fractions of $\eta=0.001$ and 0.0015 for $N=2$ and $N=3$ colloids, respectively. In the case of $N=2$, we position two colloids at a fixed distance $R$ and measure the ion-averaged PM forces ${\bf F}^{PM}_i(R)$ for a range of distances. 
The potential of mean force can then be generated from a straightforward integration over the distance, 
\begin{equation}\label{U2PM}
    U_{2}^{PM}(R) = \int_R^\infty dR' \frac{\mathbf{F}^{PM}_i(R') \cdot \mathbf{R}_{ij}}{R'},
\end{equation}
with $\mathbf{R}_{ij}={\bf R}_j-{\bf R}_i$. In this calculation we consider the (opposite) forces on both colloids, $i=1,2$, to enhance the accuracy of $U_2^{PM}(R)$. Similarly, we determine the potential of mean force for a configuration of $N = 3$ colloids, where we restrict attention to colloids in an equilateral triangle of side length $R$.\cite{Zhang2016}  We now find the three-body component $U_3^{PM}(R)$ of the potential of mean force by integrating
\begin{eqnarray}
    U_{3}^{PM}(R) &=& \frac{2}{\sqrt{3}} \int_R^\infty dR' \frac{\mathbf{F}^{PM}_i(R') \cdot \mathbf{R}_{ij} + \mathbf{F}^{PM}_i(R') \cdot \mathbf{R}_{ik}}{R'} \nonumber \\ &&- 2 U^{PM}_2(R),\label{U3PM}
\end{eqnarray}
where the factor $2/\sqrt{3}$ needs to be included to account for the triangular configuration of the colloids. We again enhance the accuracy by averaging $U_3^{PM}(R)$ of each of the three colloids.

\begin{figure}[ht]
    \centering  
    \hspace*{-0.5cm}
    \includegraphics[width=1\linewidth]{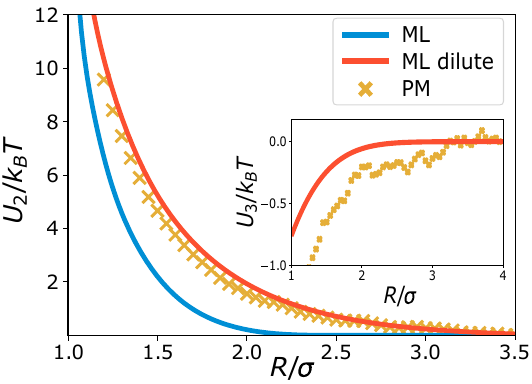}%
  \caption{The effective pair potential $U^{ML}_2(R)$ predicted by our ML models (blue and red solid lines) and the potential of mean force $U_2^{PM}(R)$ measured in primitive model (PM) simulations (yellow crosses) for a system consisting of colloids with charge $Z=90$, effective temperature $\sigma/\lambda_B = 197$,  inverse screening length $\kappa\sigma = 1.56$, and ion-to-colloid size ratio $\sigma_i/\sigma = 0.05$. Inset: effective three-body ML potential  $U^{ML}_3(R)$ and the potential of mean force $U^{PM}_3(R)$ for three colloids placed in an equilateral triangle with edge length $R$ for the same system parameters. ML models are trained on configurations with packing fractions in the range  $\eta \in \lbrack0.001,0.45\rbrack$ (blue) and on dilute configurations with $\eta \in \lbrack0.001,0.1\rbrack$ (red). }
  \label{fig:ML_PMF}
\end{figure}

In the main graph of Fig.~\ref{fig:ML_PMF}, we compare the effective ML pair potential $U^{ML}_2(R)$ of Eq.~(\ref{U2}), shown in blue, with the PM potential of Eq.~(\ref{U2PM}), represented by the yellow crosses. 
It is important to emphasize that the effective ML pair potential is trained on systems with many colloids at much higher packing fractions  ($\eta \in \lbrack 0.001,0.45\rbrack$) than the packing fraction   used to evaluate the PM potential ($\eta = 0.001$) using only 2 colloids. As a result, the effective ML pair potential may  incorporate many-body effects, which are strictly absent in the PM potential. Despite this, it remains informative to compare the two potentials. From Fig.~\ref{fig:ML_PMF}, we observe that the two potentials agree qualitatively, both representing purely pairwise repulsive interactions; however, the effective ML pair potential exhibits a significantly shorter range than the PM potential. The reason can be traced back to the highest packing fractions included in the ML training, where the ratio $y\equiv ZN/(V c_s)$ between the counterion concentration $ZN/V$ and the salt reservoir concentration $c_s$ exceeds unity. A simple estimate shows that $y=24\eta (Z\lambda_B/\sigma)/(\kappa\sigma)^2\simeq 4.5\eta$ for the system parameters considered here. In the counterion-dominated high-$\eta$ regime ($y>1$ so say $\eta>0.22$) screening becomes substantially more efficient compared to the low-$\eta$ regime, where $y<1$.  For this reason, we also retrained $U^{ML}(\{{\bf R}\})$ using only configurations with packing fractions restricted to $\eta\in[0.001,0.1]$ such that $y<1$ by a reasonably safe margin,  resulting in a total of 204 configurations. Note that the packing fraction range is still higher than the packing fraction used to evaluate the PM potential. The resulting low-$\eta$ forces are again compared with the PM forces in Fig.~\ref{fig:force_compare}(b).
 We find that the ML potential reproduces the PM forces, though less successfully than for the full potential, as indicated by the lower coefficient of determination, $R^2 = 0.865$.   
The resulting ML potential, presented as the orange curve in Fig.~\ref{fig:ML_PMF},  more closely approaches the PM pair potential under these low-density conditions. Moreover, the corresponding three-body ML potential in a triangular configuration, shown in the inset, approaches the effective PM potential, with both being purely attractive and exhibiting comparable strength and range. 
An attractive three-body contribution to the potential, as observed in simulations \cite{Lin2021} and in experiments,\cite{Dobnikar2004, Brunner2004, Merrill2009} can promote particle aggregation, with many-body interactions becoming particularly significant  in dense clusters of particles. 
Interestingly, the range of $U_3^{ML}(R)$ is again slightly shorter than that obtained from the PM simulations. This suggests that the ML training is overly influenced by the ion-averaged PM forces sampled near the upper bound of the dilute regime at $\eta=0.1$.  It is also important to note that small numerical uncertainties in $U^{ML}_2$ can lead to relatively large contributions to the effective three-body ML potential $U^{ML}_3$.

\begin{figure}[ht]
\hspace*{0cm}
    \includegraphics[width=1\linewidth]{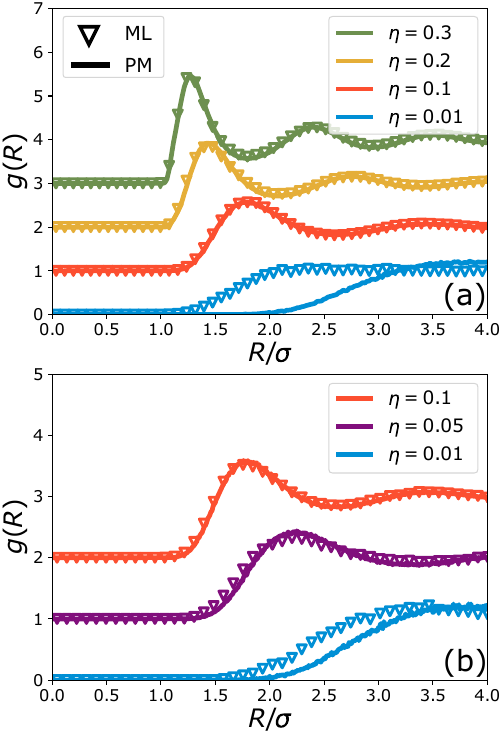}%
  \caption{Radial distribution function $g(R)$ for the colloidal dispersion described in Section \ref{sec:colsys}
  at several packing fractions $\eta$ (see labels) as obtained from primitive model (PM) simulations (lines) and from coarse-grained simulations using ML potentials  (triangles), in (a) trained on configurations with packing fractions in the range $\eta \in [0.001, 0.45]$,  with the curves shifted upwards for $\eta = 0.1$, $\eta = 0.2$, and $\eta = 0.3$ by $+1$, $+2$, and $+3$, respectively, and in (b) on dilute configurations with $\eta \in \lbrack 0.001, 0.1 \rbrack$,  with the curves shifted upwards for $\eta = 0.05$ and $\eta = 0.1$ by $+1$ and $+2$, respectively}.  \label{fig:roy_rdf_compare_combined}
\end{figure}

Next, we go beyond the properties of the two- and three-colloid systems and consider the colloid-colloid radial distribution function $g(R)$ as a function of the center-to-center distance $R$ between the colloidal particles for a variety of packing fractions. To this end, we perform molecular dynamics simulations of an effective colloids-only system using the ML potentials, and measure the radial distribution function.
In  Fig. \ref{fig:roy_rdf_compare_combined}, we plot $g(R)$ as obtained from primitive model simulations and from molecular dynamics simulations using the ML potential, in (a) trained on all configurations and in (b) only on the configurations in the dilute regime $\eta\in[0.001,0.1]$ as discussed above. Figure \ref{fig:roy_rdf_compare_combined}(a) reveals excellent agreement at the highest three packing fractions $\eta\in[0.1,0.3]$. 
 However, the  ML potential fails to reproduce the primitive model simulation results at low densities; for the dilute system at $\eta = 0.01$ the ML prediction exhibits a much shorter interaction range than the primitive model benchmark. 
Figure \ref{fig:roy_rdf_compare_combined}(b) shows agreement for $\eta=0.1$, comparable to the agreement observed in panel (a) for the full training set, and for $\eta=0.05$, the $g(r)$'s match well, although not perfectly. For $\eta=0.01$, the agreement improves significantly compared to panel (a) due to the enhanced range of the repulsion, predicted by this version of the ML potential.  Nevertheless, a significant deviation from the primitive model structure persists and perfect agreement would probably require training on an even smaller set of low $\eta$. 

We conclude that the effective ML potential for this low-salt suspension, where the Debye length of the electrolyte reservoir is on the order of the particle diameter ($\kappa^{-1}=1.54\sigma$ in this case), can effectively capture the ion-averaged forces ${\bf F}^{PM}_i$ and the effective colloids-only potential as learned from primitive model simulations for  relatively dense colloid suspensions, although  its performance becomes worse at  low densities. Even though the ML potential only depends on a particle's local environment, and does not include the long-range interactions of the PM model, it can be used to simulate a system of charged colloids, owing to charge screening in the electrolyte. However, the density dependence remains challenging to capture across the full range of packing fractions. In particular, we find that the high-$\eta$ regime, dominated by counterions, and the low-$\eta$ regime, dominated by the background electrolyte, require separate treatments. Further work is needed to quantify and fully elucidate these behaviors. 
The density dependence of the potential should be explicitly included during  ML potential training. This can be achieved by incorporating density-dependent weights in the linear regression, using neural networks trained also on density, or by including higher-body descriptors,\cite{PhysRevB.110.064101} long-range descriptors,\cite{loche2025fast} or Ewald summation techniques.\cite{kim2025universalaugmentationframeworklongrange}

\section{Salt-Free Asymmetric Electrolyte Potential}\label{sec:Z50}
In Section \ref{sec:colsys}, the investigation targeted an experimental colloidal system falling into the Poisson-Boltzmann regime, which can thus be described accurately by DLVO theory. We now turn our attention to a salt-free colloidal system at low effective temperature $\sigma/\lambda_B$ in the strong-coupling regime. This regime is characterised by strong electrostatic interactions and a quasi-two-dimensional layer of counterions condensing on the surface of the colloids rather than the diffuse three-dimensional electric double layer around the colloid, characteristic of the Poisson-Boltzmann regime.  
The salt-free colloidal dispersion, which can also be seen as an extremely asymmetric electrolyte, consists of colloids with charge valency $Z = 50$, and since the system is salt-free with reservoir screening parameter $\kappa\sigma =0$, we add $50$ charge-neutralizing counterions for every colloid, with ion-to-colloid size ratio $\sigma_i/\sigma = 0.05$. Since we focus on the strongly coupled electrostatic regime, we perform simulations at low temperatures 
and generate four separate ML potentials corresponding to $\sigma/\lambda_B = 1.2$, $\sigma/\lambda_B = 1.6$, $\sigma/\lambda_B = 2.0$, and $\sigma/\lambda_B = 2.4$.

We train the machine-learned potentials on ion-averaged forces, generated via primitive model simulations of $N = 64$ colloids, for 55 packing fractions in the range $\eta \in [0.001, 0.45]$, where we collect four configurations $\{\mathbf{R}\}$ per packing fraction for a total of 220 configurations per potential. We used the techniques described in Section \ref{sec:traingen} to perform the PM simulation and generated ML potentials with the framework described in Section \ref{sec:ML-procedure}. All ML potentials generated in this section are composed of $D = 20$ symmetry functions.

\begin{figure}[ht]
    \centering  
    \hspace*{-0.5cm}
    \includegraphics[width=1.0\linewidth]{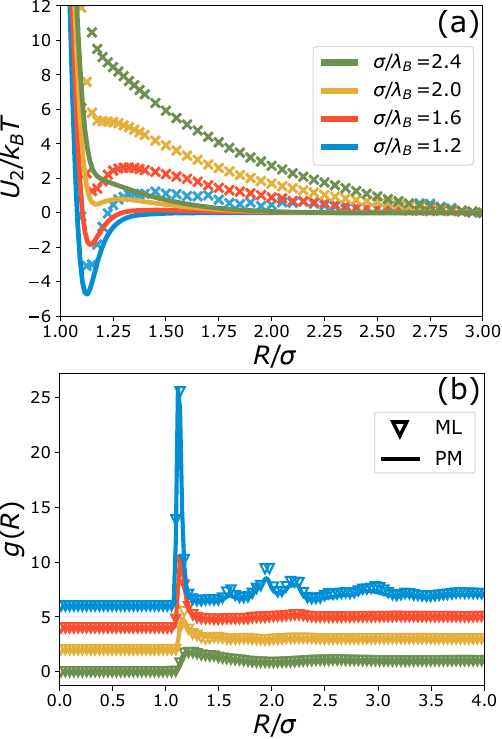}%
  \caption{(a) The potential of mean force $U^{PM}_2(R)$ (crosses) and the potential $U^{ML}_2(R)$ predicted by our ML model (solid lines), for a system consisting of colloids with charge valency $Z=50$ and counterions, with ion-to-colloid size ratio $\sigma_i/\sigma = 0.05$, and effective temperatures $\sigma/\lambda_B \in [1.2, 2.4]$.
  The ML potential is trained on primitive model simulation of $N=32$ colloids with packing fractions in the range  $\eta \in \lbrack0.001,0.45\rbrack$ and may incorporate many-body effects, whereas the PM potential is strictly a two-body potential, since it was determined using simulations of two isolated colloids, albeit density-dependent in this salt-free case.
  (b) Radial distribution function $g(R)$ for the salt-free colloidal dispersion described above, at packing fraction $\eta = 0.2$ as obtained from primitive model simulations (lines) and from coarse-grained simulations using ML potentials (triangles).  The curves for $\sigma/\lambda_B = 2.0$, $\sigma/\lambda_B = 1.6$, $\sigma/\lambda_B = 1.2$ are shifted upwards by $+2$, $+4$, and $+6$,  respectively.}
  \label{fig:Z50_overview}
\end{figure}

In Fig. \ref{fig:Z50_overview}(a), we compare the effective  ML pair potential $U_2^{ML}(R)$ (solid lines) with the potential of mean force $U_2^{PM}(R)$ (crosses) obtained from primitive model simulations for various values of $\sigma/\lambda_B$. The potential of mean force $U_2^{PM}(R)$ is determined using primitive model simulations with two colloids and $100$  counterions in a cubic simulation box with side length $L = 6\sigma$, corresponding to a colloid packing fraction $\eta=0.0048$. In these simulations, the colloids are held fixed at different separations while the forces acting on them are thermally averaged over the counterion configurations. The PM potential $U_2^{PM}(R)$ is then calculated using  Eq.~(\ref{U2PM}).
From both potentials in Fig. \ref{fig:Z50_overview}(a), we observe features characteristic of interactions in the strong coupling regime. At high temperatures, $\sigma/\lambda_B = 2.4$, the two-body potentials are purely repulsive. As the temperature decreases, an attractive well starts to form, until this well becomes the dominant feature at $\sigma/\lambda_B = 1.2$. Surprisingly, the PM potential and the ML potential show qualitative agreement  at $\sigma/\lambda_B = 1.2$, as the attractive well appears at the same temperature and with comparable depth in both cases. In this regime, the counterions are condensed to the surface of the colloids, which explains the reasonable match between the effective ML potential and the PM potential. 
However, quantitative differences develop at higher values of  $\sigma/\lambda_B$
--- as expected since in the salt-free  case the potential of mean force only reaches its dilute limit---bare colloid-colloid Coulomb interactions---at extremely low densities. In fact we estimate on the basis of Ref.\citenum{dijkstra1998vapour} that the potential of mean force remains density-dependent upon dilution  down to $\eta \simeq 10^{-15}$.
Furthermore, we note that the qualitative differences between ML  and PM potentials may also arise from many-body effects that are included in  the effective ML potentials  as they are trained on systems with many colloids at relatively high densities and thus high counterion concentrations. 
However, the ML potential does capture the qualitative change in colloid interactions with decreasing temperature, showing a reduction in  repulsion and the emergence of   a short-ranged attractive well.

In Fig. \ref{fig:Z50_overview}(b), we compare the colloid-colloid radial distribution function $g(R)$ obtained from primitive model simulations (solid lines) with those generated using the  ML potential (triangles)  at a packing fraction of $\eta = 0.2$. The two $g(R)$'s agree for a range of temperatures, and both predict structural changes with changing temperature. At $\sigma/\lambda_B = 1.2$, the $g(R)$ indicates crystallization of the colloids, while at  $\sigma/\lambda_B = 1.6$, this crystalline structure is replaced by a homogeneous fluid structure. Upon further increasing the temperature, the peak in the $g(R)$ at $R = 1.2 \sigma$ becomes less pronounced as the potential well at this distance diminishes.
Our findings demonstrate that machine-learned potentials are effective for predicting the structure and phase behavior of salt-free colloidal suspensions in the strong coupling regime.

\section{1:1 Electrolyte Potential}\label{sec:elsys}

We have applied the ML framework to generate effective potentials for colloidal systems by coarse-graining out the ions in the suspension and determining the effective interactions between colloids surrounded by an electric double layer or a two-dimensional layer of condensed counterions.
To test the limits of the ML framework, we now apply it to a fundamentally different system, by generating effective potentials for cation-cation interactions in a 1:1 electrolyte. This relatively simple system, consisting of cations and anions in a symmetric two-component mixture, lacks the charge and size asymmetries present in the three-component, asymmetric colloidal system discussed in Section   \ref{sec:colsys}.
The parameters are chosen such that we model an aqueous 1:1 electrolyte at room temperature, where we set the Bjerrum length to $\lambda_B = 0.7 \ \text{nm}$ and the ion diameter to $\sigma_i = 0.35 \ \text{nm}$ for both the cations and anions. This corresponds to an effective temperature of $\sigma_i/\lambda_B = 0.5$. Since the critical point of a 1:1 electrolyte lies around $\sigma_i/\lambda_B = 0.05$,\cite{RomeroEnrique2000, elec_Hynninen2005}  we do not expect gas-liquid phase separation to play a role in our simulations. 

We generate the training data using primitive model simulations with $N_+ = 64$ cations and $N_- = 64$ anions over a range of packing fractions $\eta \in [0.001, 0.1]$, where the packing fraction is defined as $\eta = (N_+ + N_-)\pi \sigma_i^3/(6V)$. This corresponds to salt concentrations in the range $c_s \in [0.074, 7.4] \ \text{M}$. In total, we generate 218 configurations,  using the positions and the averaged forces acting on the cations as input for the ML potential. We use the same techniques as described in Section \ref{sec:ML-procedure}, with the cations now playing the role of the colloids, and generate ML potentials with a cut-off at $R_c = 4.0 \sigma_i$.

\begin{figure}[ht]
    \centering  
    \hspace*{-0.5cm}
    \includegraphics[width=1\linewidth]{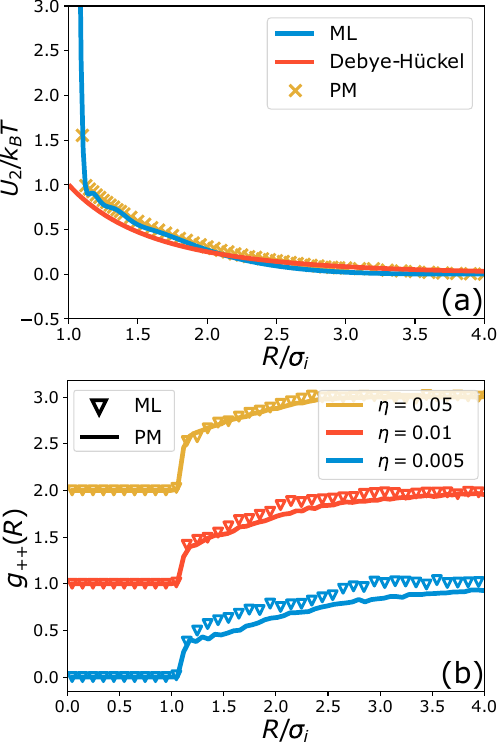}%
  \caption{(a) The effective cation-cation pair potential $U_2(R)$ as predicted by our ML models (blue solid lines), according to the Debye-H\"{u}ckel Theory $U^{DH} = \lambda_B \exp{[-\kappa R]}/R$ (red solid line), and measured in primitive model (PM) simulations (yellow crosses) for a 1:1 electrolyte at effective temperature $\sigma_i/\lambda_B = 0.5$ and $\eta=0.01$. The ML potential is trained on configurations with packing fractions in the range $\eta \in \lbrack0.001,0.1\rbrack$. (b) Cation-cation radial distribution function $g_{++}(R)$ in an electrolyte at several packing fractions $\eta$ (see labels) as obtained from primitive model simulations (lines) and from coarse-grained simulations using ML potential (triangles).  The curves for $\eta = 0.01$ and $\eta = 0.05$ are shifted upwards by $+1$ and $+2$, respectively.} 
  \label{fig:elec_ML_PMF}
\end{figure}

As in Section \ref{sec:colsys}, we compare the machine-learned two-body potential $U_2^{ML}(R_{})$, composed of $D = 20$ symmetry functions,  to a potential of mean force $U_2^{PM}(R)$ in Fig. \ref{fig:elec_ML_PMF}(a). The potential of mean force is computed using primitive model simulations at a packing fraction $\eta = 0.01$, corresponding to a salt concentration of $c_s = 0.74 \text{M}$, using Eq.~(\ref{U2PM}). In Fig. \ref{fig:elec_ML_PMF}(a), we observe excellent agreement between the primitive model potential $U_2^{PM}(R)$, indicated with yellow triangles, and the ML potential $U_2^{ML}(R)$, the blue line, since the two curves almost completely overlap. This agreement between the ML potential and the PM potential is even better than with the Debye-H\"{u}ckel potential, $U_2^{DH}(R) = \lambda_B \exp{[-\kappa R]}/R$, commonly used for describing effective ionic interactions. The ML two-body potential accurately captures both the screened repulsive electrostatic interactions and the hard-core repulsion. For these parameters, the three-body contribution is insignificant and is therefore not shown.

The agreement between the ML potential and the PM potential also manifests in the structures generated by simulations. In Fig. \ref{fig:elec_ML_PMF}(b), we compare the cation-cation radial distribution function $g_{++}(R)$ obtained from primitive model simulations (solid lines) with that from ML simulations (triangles). At the highest packing fraction considered, $\eta = 0.05$, the two $g_{++}(R)$'s agree well, indicating that the ML potential can reliably capture the structural behavior of the electrolyte under these conditions.  However, at lower packing fractions, the agreement decreases, similar to our findings in Section \ref{sec:colsys}, where we generated effective ML potentials for charged colloids. For $\eta = 0.005$, the $g_{++}(R)$ obtained using the ML potential is higher at small cation separations than that from primitive model simulations, indicating that the ML potential slightly underestimates the repulsion at lower concentrations. This discrepancy arises because  high-density configurations with relatively strong forces dominate  the training data, leading to reduced performance of the ML potential at lower concentrations.

Surprisingly, the ML framework is not only suitable for reproducing the interactions between colloidal particles, for which it was originally developed, but also for capturing interactions between ions in an electrolyte. In contrast with the colloidal system, the particles that are coarse-grained out are of equal size and carry charges of equal magnitude as the particles for which the effective potential is determined. Nevertheless, this symmetry does not hinder the successful generation of effective potentials. However, the long-range nature of the Coulomb interactions presents a challenge for the ML framework, since at low densities, where long-range interactions are most significant, the ML potential fails at perfectly replicating the primitive model results.

\section{Conclusion}

In this work, we have applied a machine-learning framework to generate potentials that mimic effective interactions of charged particles, either colloids or ions, suspended in an electrolyte. We considered colloid-colloid, colloid-ion, and ion-ion interactions within the primitive model, where colloidal and ionic particles are treated explicitly,  and the solvent is represented as a dielectric continuum. Effective like-charge attractions between colloidal particles can arise in primitive model simulations when the colloidal surface charge density is high and the effective temperature is low, indicative of the strongly coupled electrostatic regime. However,  large-scale primitive model simulations of colloidal dispersions require the explicit treatment of a large number of ions and the evaluation of long-range electrostatic interactions, making these simulations computationally demanding.  This motivated us to generate machine-learned potentials that include many-body components, enabling us to perform significantly faster simulations while retaining the essential physics of the system. 

The machine-learned potentials were first tested on a system of micron-sized colloids suspended in an electrolyte, using parameters matching an experimental system.\cite{Royall2006}  The ML potential accurately reproduced the structures found in primitive model simulations across most densities, although it failed to replicate the primitive model results at low packing fractions of $\eta < 0.01$. Interestingly, the ML potential exhibited an attractive three-body component that aligned with the attractive interactions found in primitive model simulations.
The second system on which the ML framework was tested was a highly asymmetric electrolyte, consisting of colloids of charge valency $Z = 50$ and monovalent counterions. In this system, the like-charge attraction between colloids, characteristic of the strong-coupling regime, was well-captured by the ML potentials. These attractive interactions had a pronounced effect on the structure of the system, resulting in colloid crystallization upon reducing the effective temperature $\sigma/\lambda_B$ at fixed packing fraction.
Finally, the ML framework was tested on the extreme case of an aqueous 1:1 electrolyte, where the effective cation-cation interaction potential in the presence of anions was successfully reproduced. Surprisingly, even in this peculiar case, the structural behavior of the electrolyte observed in the primitive simulations was accurately captured  by the ML potential simulations of the effective cation-only system. 


 The ML framework developed here  enables large-scale simulations of charged particles, allowing us to probe  the phase behavior of highly charged colloidal dispersions, even at low or zero salt. 
However, the current approach is limited by the need to generate a separate ML potential for each temperature and by its poor representation of low colloid densities. These limitations could be addressed in future work by incorporating temperature- and density-dependent weights in the linear regression, or by using more advanced techniques such as neural networks, higher-body descriptors,\cite{PhysRevB.110.064101} long-range descriptors,\cite{loche2025fast} or Ewald summation methods.\cite{kim2025universalaugmentationframeworklongrange} 
Nevertheless, the framework is still instrumental for exploring several aspects of how salt affects phase behavior, as both  experiments \cite{Larsen1997, Gomez2009, Wang2024} and simulations \cite{Hynninen2009} have shown that finite salt concentrations can change the effective colloid-colloid  interactions in unexpected ways. 
For highly charged colloids, the effect of adding salt can also be studied by training ML potentials across a range of screening lengths, corresponding to various salt concentrations, at physically relevant coupling parameters. By comparing these potentials, it becomes possible to study like-charge attractions within the many-body interactions, analyze their dependence on screening length, and determine whether these attractive terms can drive phase coexistence.
 In future work, we will examine how attractive three-body interactions and added salt influence the phase behavior of charged colloids.
Finally, we note that interfaces between coexisting phases are as intriguing as they are challenging, since the density-dependent two-body repulsions and three-body attractions identified for certain parameter regimes were derived from training on quasi-homogeneous rather than strongly heterogeneous states. Assessing to what extent such density-dependent ML potentials can capture  interfacial phenomena in phase-separated colloidal dispersions remains an open and challenging problem. 
\begin{acknowledgments}

The authors thank Floris van den Bosch for his contributions to an early version of this project. The authors also thank Tim Veenstra for many useful discussions. T.t.R. and M.D. acknowledge funding from the European Research Council (ERC) under the European Union's Horizon 2020 research and innovation programme (Grant agreement No. ERC-2019-ADG 884902 SoftML).
\end{acknowledgments}

\section*{Conflict of Interest}
The authors have to conflicts to disclose.

\section*{Author Contribution}
\textbf{Thijs ter Rele}: Conceptualization (equal); Data Curation (lead); Formal Analysis (lead); Investigation (lead); Methodology (equal); Software (lead); Visualization (lead); Writing - original draft (lead); Writing - review \& editing (equal). 
\textbf{Gerardo Campos-Villalobos}: Conceptualization (equal);  Investigation (supporting);  Methodology (equal); Software (supporting); Visualization (supporting); Writing - review \& editing (supporting). 
\textbf{Ren\'{e} van Roij}: Conceptualization (equal);  Investigation (supporting); Methodology (equal); Visualization (supporting);   Writing - review \& editing (equal). 
\textbf{Marjolein Dijkstra}: Conceptualization (equal), Funding acquisition (lead); Investigation (supporting); Methodology (equal); Supervision (lead);  Visualization (supporting);  Writing - review \& editing (equal).

\section*{Data Availability Statement}

The data that support the findings of this study are available from the corresponding author upon reasonable request.

\appendix

\section{Symmetry Functions Gradients}\label{sec:symgrad}
The forces on the colloids in ML potential simulations depend on the gradient of the symmetry function $G(i)$ with respect to particle $j$. You have to consider if the gradient is taken with respect to the same particle that belongs to the symmetry function that is being determined ($\nabla_i G^{(2)}(i)$) or with respect to some other particle ($\nabla_j G^{(2)}(i)$), since this has an influence on the gradient. For the two-body symmetry function, the gradients are
\begin{align}
\hspace{-1.cm}
    &\nabla_i G^{(2)}(i) = \\ \nonumber
    &\sum_{i\neq j}e^{-\gamma(R_{ij} - R_s)^2 } \left[\frac{d f_c (R_{ij})}{dR_{ij}} - 2\gamma (R_{ij} - R_s)\right] \frac{\mathbf{R}_{ij}}{R_{ij}}
\end{align}
and 
\begin{align}
\hspace{-1.cm}
    &\nabla_j G^{(2)}(i) = \\ \nonumber
    &e^{-\gamma(R_{ij} - R_s)^2 } \left[\frac{d f_c (R_{ij})}{dR_{ij}} - 2\gamma (R_{ij} - R_s)\right] \frac{\mathbf{R}_{ij}}{R_{ij}},
\end{align}
where we use $\mathbf{R}_{ij} = \mathbf{r}_i - \mathbf{r}_j$ and $R_{ij} = |\mathbf{R}_{ij}|$. For the three-body symmetry functions the gradients are
\begin{align}
     \nabla_i G^{(3)}(i) = & 2^{1-\xi}\sum_{i,k\neq j} \Omega(\mathbf{R}_{ij}, \mathbf{R}_{ik}) \nonumber
     \\ &\Big[\left(\phi(\mathbf{R}_{ij}, \mathbf{R}_{ik}) - 2\gamma + \chi(R_{ij}) \right) \mathbf{R}_{ij} \nonumber
     \\ &+ \left(\phi(\mathbf{R}_{ik}, \mathbf{R}_{ij}) - 2\gamma + \chi(R_{ik}) \right)\mathbf{R}_{ik} \Big]
\end{align}
and 
\begin{align}
     \nabla_jG^{(3)}(i) = & 2^{1-\xi} \sum_{k\neq i, k\neq j} \Omega(\mathbf{R}_{ij}, \mathbf{R}_{ik}) \nonumber
     \\ &\Big[ - \left(\phi(\mathbf{R}_{ij}, \mathbf{R}_{ik}) - 2\gamma + \chi(R_{ij}) \right) \mathbf{R}_{ij} \nonumber
     \\ &+ \left(\psi(\mathbf{R}_{ij}, \mathbf{R}_{ik}) - 2\gamma + \chi(R_{jk}) \right)\mathbf{R}_{jk} \Big].
\end{align}
These gradients contain the following functions
\begin{align}
    \Omega(\mathbf{R}_{ij}, \mathbf{R}_{ik}) &= \left(1 + \lambda \cos{\theta(\mathbf{R}_{ij}, \mathbf{R}_{ik})}\right)^\xi e^{-\gamma\left(R_{ij}^2 +R_{ik}^2 + R_{jk}^2\right)} \nonumber
    \\& \ \ \ \ f_c(R_{ij}) f_c(R_{ik}) f_c(R_{jk})
    \\ \psi(\mathbf{R}_{ij} , \mathbf{R}_{ik}) &= -\frac{1}{R_{ij}R_{ik}} \frac{\lambda \xi}{1 + \lambda \cos{\theta(\mathbf{R}_{ij}, \mathbf{R}_{ik})}}
    \\ \phi(\mathbf{R}_{ij} , \mathbf{R}_{ik}) &= \frac{1}{R_{ij}}\left[\frac{1}{R_{ik}} - \frac{1}{R_{ij}}\cos{\theta(\mathbf{R}_{ij}, \mathbf{R}_{ik})} \right] \nonumber
    \\ & \ \ \ \ \frac{\lambda \xi}{1 + \lambda \cos{\theta(\mathbf{R}_{ij}, \mathbf{R}_{ik})}}
    \\ \chi(R) &= \frac{1}{R f_c(R)} \frac{d f_c(R)}{d R}. 
\end{align}

\noindent For $G^{(2)}$: \\
$\gamma \in \{0.01, 0.1, 1, 2, 4, 8, 16\} \sigma^{-2}$\\ 
$R_s \in \{0.0, 0.1, 0.2, 0.3, 0.4, 0.5, 0.6, 0.7, 0.8, 0.9, 1.0\} \sigma$. \\ \\
\noindent For $G^{(3)}$:\\
$\gamma \in \{0.01, 0.1, 1, 2, 4, 8, 16\} \sigma^{-2}$\\
$\lambda \in \{1, 2, 4, 8, 16, 32\}$\\
$\xi \in \{1, -1\}$

\bibliography{bibfile}

\end{document}